\documentclass[preprint,aps,showpacs]{revtex4}       
\usepackage{graphicx}
\newcommand{\be}{\begin{equation}}
\newcommand{\ee}{\end{equation}}
\newcommand{\bea}{\begin{eqnarray}}
\newcommand{\eea}{\end{eqnarray}}
\newcommand{\Tr}{{\rm Tr}}
\newcommand{\cL}{{\cal L}}
\begin{document}
\title{Deformed Nuclei in a Chiral Model}
\author{S. Schramm}
\email{schramm@theory.phy.anl.gov} \affiliation{Argonne National
Laboratory, 9700 S. Cass Avenue, Argonne IL 60439, USA}
\date{\today}

\begin{abstract}
We investigate the deformation properties of atomic nuclei in a
hadronic chiral SU$_{\rm flavor}$(3) model approach. The
parameters are fitted to hadron mass properties and adjustments
for spherical finite nuclei have been performed. Using these
parameters the deformation of a series of light and heavy nuclei
are obtained in a two-dimensional self-consistent calculation. In
addition a case of superdeformation in a heavy nucleus is studied.
\end{abstract}
\pacs{21.60.-n,12.40.-y} \maketitle

\section{Introduction}

Relativistic descriptions of strong interaction physics using
hadronic degrees of freedom have again become more popular in
recent years.
One of the areas where hadronic models have been studied
is the regime of highly excited hot and dense nuclear matter as explored
experimentally using ultrarelativistic heavy-ion collisions.
Towards the phase transition to a chirally restored
phase a variety of models loosely connected to the chiral $\sigma$ model have
been investigated. Extensions including vector meson
fields ($\sigma-\omega$ model) and the strangeness degree
of freedom by incorporating hyperons and strange mesons
have been discussed. In that context density and temperature dependent hadronic masses
and the structure of the phase transition were calculated. In general, however, those models fail in
their extrapolation back to normal matter and lack an
adequate description of the groundstate properties of nuclear matter, especially
they generate too large values for its compressibility.

On the other hand there has been significant progress in the
description of finite nuclei using a relativistic meson field
(RMF, ``Walecka'' type) ansatz for modelling the basic degrees of freedom of
the system \cite{Walecka,Bodmer}. Those calculations are quite successful in describing
nuclear properties over a wide range of mass numbers \cite{Reinhard,Ring1}.

It seems natural to study model descriptions that combine both of those regimes.
There has been some effort in the past to follow this
path. It turned out that one can get a sensible behavior of highly
excited matter, a good description of nuclear saturation as well as
a reasonable description of nuclei and hypernuclei with a single model
and a single set of parameters.

In this article we follow and extend this approach.
We first determine an improved parameter set \cite{beckmann}
in a fit to spherical nuclei. We then expand our calculations to two-dimensional
axially symmetric systems. We study several test cases of deformed and superdeformed nuclei
and perform a general survey of the nuclear quadrupole deformation across the whole nuclear chart.
We conclude with a critical analysis of necessary improvements to the current approach.

\section{Model Description}
In our calculation we use a hadronic model based on a chiral SU(3) ansatz in
a non-linear realization of chiral symmetry.
The basic hadron fields are the SU(3) multiplets for the baryons and mesons,
respectively, including the baryon octet $B$ that contains nucleons and the $\Lambda, \Sigma$
and $\Xi$ hyperons, and the corresponding octets and singlets for scalar, pseudoscalar, vector and
axialvector mesons. In the non-linear realization the pseudoscalar degrees of freedom generate
chiral rotations of the baryons:
\be
  B ~~=~~ u^\dagger \Psi_L u \qquad = u \Psi_R u^\dagger    .
\ee
with $u = \exp(\pi^a\lambda^a/\sigma_o)$.
$\psi_L$ and $\psi_R$ are the left-/righthanded multiplets in the linear representation
of the theory.
The 3x3 matrix $u$ contains the pseudoscalar fields $\pi^a$\cite{paper3}
\be
\label{psmatrix}
\frac{1}{\sqrt{2}}\pi_a \lambda^a
=\left (\begin {array}{ccc}
  \frac{1}{\sqrt{2}}\left ( \pi^0+{\frac {\eta^8}{\sqrt {1+2\,{w}^{2}}}}\right )&\pi^{+}
 &2\,{\frac {K^{+}}{w+1}}\\
  \noalign{\medskip}\pi^{-}&\frac{1}{\sqrt{2}}\left
 (- \pi^0+
 {\frac {\eta^8}{\sqrt {1+2\,{w}^{2}}}}\right )&2\,{\frac { K^0
  }{w+1}}\\\noalign{\medskip}2\,{\frac {K^-}{w+1}}&2\,{\frac {
 \bar{K}^0}{w+1}}&-{\frac {\eta^8\,\sqrt {2}}{\sqrt {1+2\,{w}^{2}}}}
 \end {array}\right ) .
\ee
The values $\sigma_0$ and $\zeta_0$ are the vacuum expectation values of the scalar meson fields
as discussed below.
The renormalization factors containing the ratio
$w=\sqrt{2}\zeta_0/\sigma_0$ are included
to obtain  the canonical form
of the kinetic energy terms for the pseudoscalar mesons \cite{paper3}.
Note that after the transformation (1) the resulting baryons $B$ transform
vectorially.

The baryons couple to the scalar and vector mesons (3$x$3 matrices $W$) according to the general
SU(3) scheme
\be
\label{FandD}
{\cal L}_{\mbox{BW}} =
-\sqrt{2}g_8^W \left(\alpha_W[\bar{B}{\cal O}BW]_F+ (1-\alpha_W)
[\bar{B} {\cal O}B W]_D \right)
- g_1^W \frac{1}{\sqrt{3}} \Tr(\bar{B}{\cal O} B)\Tr W  \, ,
\ee
with $[\bar{B}{\cal O}BW]_F:=\Tr(\bar{B}{\cal O}WB-\bar{B}{\cal O}BW)$ and
$[\bar{B}{\cal O}BW]_D:= \Tr(\bar{B}{\cal O}WB+\bar{B}{\cal O}BW) - \frac{2}{3}
\Tr (\bar{B}{\cal O} B) \Tr W$ and ${\cal O} = \left\{1\!\!1, \gamma_{\mu}\right\}$ .
For the parametrization discussed below we get a ratio of the $D$ (anticommutator)
and $F$-type coupling (commutator) of about $D/F \sim 0.34$. In the case of the vector
mesons in accordance with vector meson dominance arguments
we restrict ourselves to pure $F$ coupling. This means that in contrast to the scalar case
the neutral vector meson made of strange quarks, the $\phi \sim (\bar{s}s)$ does
not couple to nucleons and therefore does not enter
the equations for nuclei without hyperons. It would be, however,
phenomenologically consistent to add a small $D$ coupling admixture to the baryon-vector meson
coupling. The effect of such an admixture should be studied in future investigations.
The self-interactions of the scalar mesons are taken into account up to 4th order in the fields.
In addition the interaction terms with the glueball field $\chi$
generate a logarithmic potential. The field $\chi$ mimics the breaking of the QCD scale invariance
due to the gluon condensate $\chi \sim <F^a_{\mu\nu} F^{a,\mu\nu}>$. The specific structure
of the interaction terms are a direct result of this analogy (for a more extended discussion, see
\cite{pano}).

For the study of nuclear matter and
static nuclei in a mean field approximation the following fields have to be taken
into account: The proton and neutron, the scalar meson fields
$\sigma \sim (\bar{u}u +\bar{d}d)$,
$\zeta \sim (\bar{s}s)$, the isovector
$\delta \sim (\bar{u}u -\bar{d}d)$, the scalar glueball field $\chi$,
and the time components of the isoscalar and isovector vector mesons $\omega_0$, $\rho^0_0$
and the Coulomb field $A_0$.
Restricting ourselves to those degrees of freedom the general $SU(3)$ Lagrangian \cite{pano}
reduces to the following structure. The interaction of the baryons and mesons (and the photon)
reads
\be
\cL_{\mathrm int} = -\sum\limits_i\bar{B_i}
\left[g_{i_\omega}\omega_0 \gamma_0 +g_{i\rho}\tau_3\rho_0^0 \gamma_0
+\frac{1}{2}e(1+\tau_3)A_0 \gamma_0 +m_i^* \right] B_i
\ee
where now $B$ is reduced to the isospinor $\left( ^p_n \right)$.
The various coupling constants of mesons and baryons result from the $SU(3)$ structure
(\ref{FandD}). The vector meson mass and self-interaction terms are \cite{paper3}
\be
\label{Lvec}
\cL_{\mathrm vec} = -\frac{1}{2}k_0 \frac{\chi^2}{\chi_0^2}\left(
m_\omega^2\omega_0^2+m_\rho^2\rho^2\right)
+  g_4^4\left(\omega_0^4+6\omega_0^2(\rho_0^0)^2+(\rho_0^0)^4\right)
\ee
and the interaction of the scalar fields follows as
\bea
\cL_0^{\mathrm{chi}}  &=&  -\frac{1}{2}k_0\chi^2(\sigma^2+\zeta^2+\delta^2)+
k_1(\sigma^2+\zeta^2+\delta^2)^2
+k_2\left(\frac{\sigma^4}{2}+\frac{\delta^4}{2}+3\sigma^2\delta^2+\zeta^4\right)\nonumber\\
&& + k_3\chi\sigma^2\zeta
-k_4\chi^4
 -\frac{1}{4}\chi^4\mbox{ln}\frac{\chi^4}{\chi_0^4}
+\epsilon\chi^4\mbox{ln}\frac{(\sigma^2-\delta^2)\zeta}{\sigma_0^2\zeta_0}
\eea
In addition, the terms
\be
\cL_{\mathrm ESB} =
-\left(\frac{\chi}{\chi_0}\right)^2\left[x\sigma+y\zeta\right]
\ee
break the chiral SU(3) symmetry explicitly with the coupling strengths $x=m_\pi^2 f_\pi$ and $y=\sqrt{2}m_{\mathrm K}^2 f_{\mathrm K}
-\frac{1}{\sqrt{2}}m_\pi^2 f_\pi$, where $f_\pi$ and $f_K$ are the pion and kaon decay constant,
respectively.
The baryons attain a dynamically generated mass due to their interaction with the
scalar fields. This mass generally changes in the nuclear medium. The baryon masses are given by
\be m_i^\ast =g_{i\sigma}\sigma+g_{i\delta}\delta+g_{i\zeta}\zeta\quad . \ee
The asterisk indicates that the masses in the medium shift with
the changing isoscalar scalar fields $\sigma$ and $\zeta$ and the isovector scalar field $\delta$.
Due to the general SU(3)-symmetric structure the $\delta$ meson automatically enters
the field equations for the nuclei including contributions to the nonlinear terms of the scalar
interaction. Note that in \cite{beckmann} for simplicity the $\delta$ meson was neglected although this was not
quite consistent with the general model approach as outlined here.

\section{Calculation of Nuclei}
We follow the general approach outlined in \cite{PC}.
In contrast to our previous parameter search we used a more extended set of nuclei to fit
the model parameters as shown in Table~\ref{table1}.
\begin{table}
\caption{\label{table1}Set of spherical nuclei used to fit the model parameters
of the parameter set $\chi_m$. The deviations of binding energy, charge radius, and LS splitting
values (averaged over proton,neutron) are shown.}
\begin{ruledtabular}
\begin{tabular}{llllll}
& $\delta\epsilon (\%)$ & $\delta r_{ch}(\%)$&
& $\delta\epsilon (\%)$ & $\delta r_{ch}(\%)$\\
$^{48}$Ca & 0.64 & -1.38& $^{124}$Sn & 0.24 &-0.23\\
$^{56}$Ni & -0.61& -0.53& $^{132}$Sn & -0.07 & \\
$^{58}$Ni & -0.57& -0.47& $^{136}$Sn & -0.17& \\
$^{80}$Zr & 0.81 & & $^{136}$Xe & 0.22& \\
$^{90}$Zr & 0.30 & -0.50& $^{144}$Sm & 0.31& \\
$^{84}$Se & 0.40 & & $^{190}$Pb & -0.06& -0.96\\
$^{88}$Sr & 0.29& -0.64& $^{202}$Pb & -0.18& 0.56\\
$^{96}$Pd & 0.12& &$^{208}$Pb & -0.11 & 0.46\\
$^{100}$Sn & 0.02& &$^{214}$Pb & -0.17& 0.61\\
$^{112}$Sn & 0.20& -0.02& $^{210}$Po & 0.02 & 1.04 \\
$^{120}$Sn & 0.33& &$^{214}$Ra & 0.29 &
\end{tabular}
\end{ruledtabular}
\end{table}
We only used the binding energies of the nuclei
for the $\chi^2$ fit. The charge radii, which generally show reasonable behavior,
and the LS splitting (see Table \ref{tableLS})
were obtained without including them in the fit. Comparing the errors with
similar studies in relativistic mean field theories \cite{PC} shows that
this approach can obtain the same fit quality as well-established RMF parameter sets.
\begin{table}
\caption{\label{tableLS}LS splitting for parameter set $\chi_m$. The deviations from
experiment are quoted. In brackets the corresponding
single particle states are stated.}
\begin{ruledtabular}
\begin{tabular}{lll}
& $\delta$(Ls)$_p$~(\%) & $\delta$(LS)$_n$~(\%)\\
$^{16}$O &-13.7 (1p)& -9.8 (1p)\\
$^{40}$Ca & -24.1 (1d) & -12.4 (1d)\\
$^{132}$Sn & -3.3 (2d)& -12.4 (2d)\\
$^{208}$Pb & 0.8 (2d)& -29.7 (3p)
\end{tabular}
\end{ruledtabular}
\end{table}
The fit parameters for the fit $\chi_m$ is shown in table \ref{table2}.
Using the fitted parameters we check that the description of vacuum masses and nuclear matter
values are still in good agreement with experiment. The result is listed in Table~\ref{table4}. Hadron masses changed only slightly.
The nuclear compressibility $\kappa$ has a value of
\begin{equation}
\kappa \equiv 9 \rho^2 \frac{\partial^2(E/A)}{\partial \rho ^2} = 215.33~ {\rm MeV}~~.
\end{equation}
Note that the model produces a satisfactory value for the asymmetry energy defined as
\begin{equation}
E_{asym} = \frac{1}{2} \rho_o^2 {\partial^2(E/A) \over
\left[\partial (\rho_p-\rho_n)\right]^2 }= 31.9~ {\rm MeV}~~,
\end{equation}
which is more in accordance with phenomenological values as compared to the usually too large
values for $E_{asym}$ in RMF calculations \cite{Rutz}.
\begin{figure}
\includegraphics[width = 7cm]{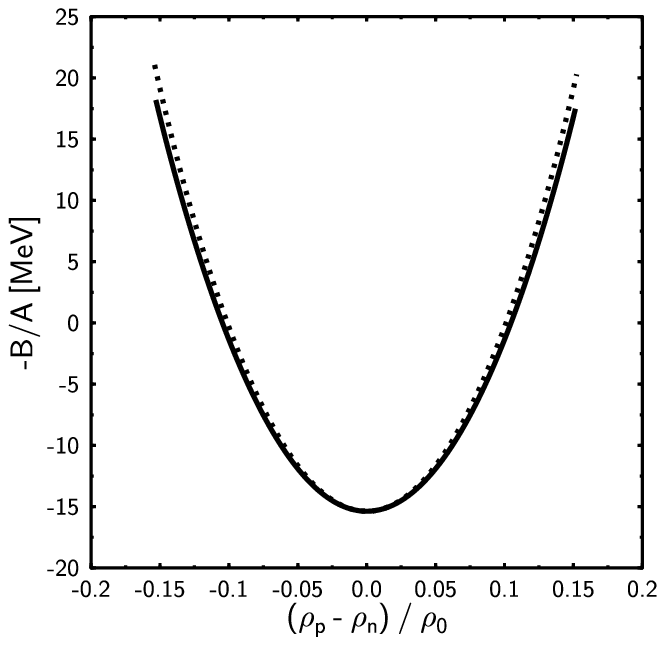}
\includegraphics[width = 7cm]{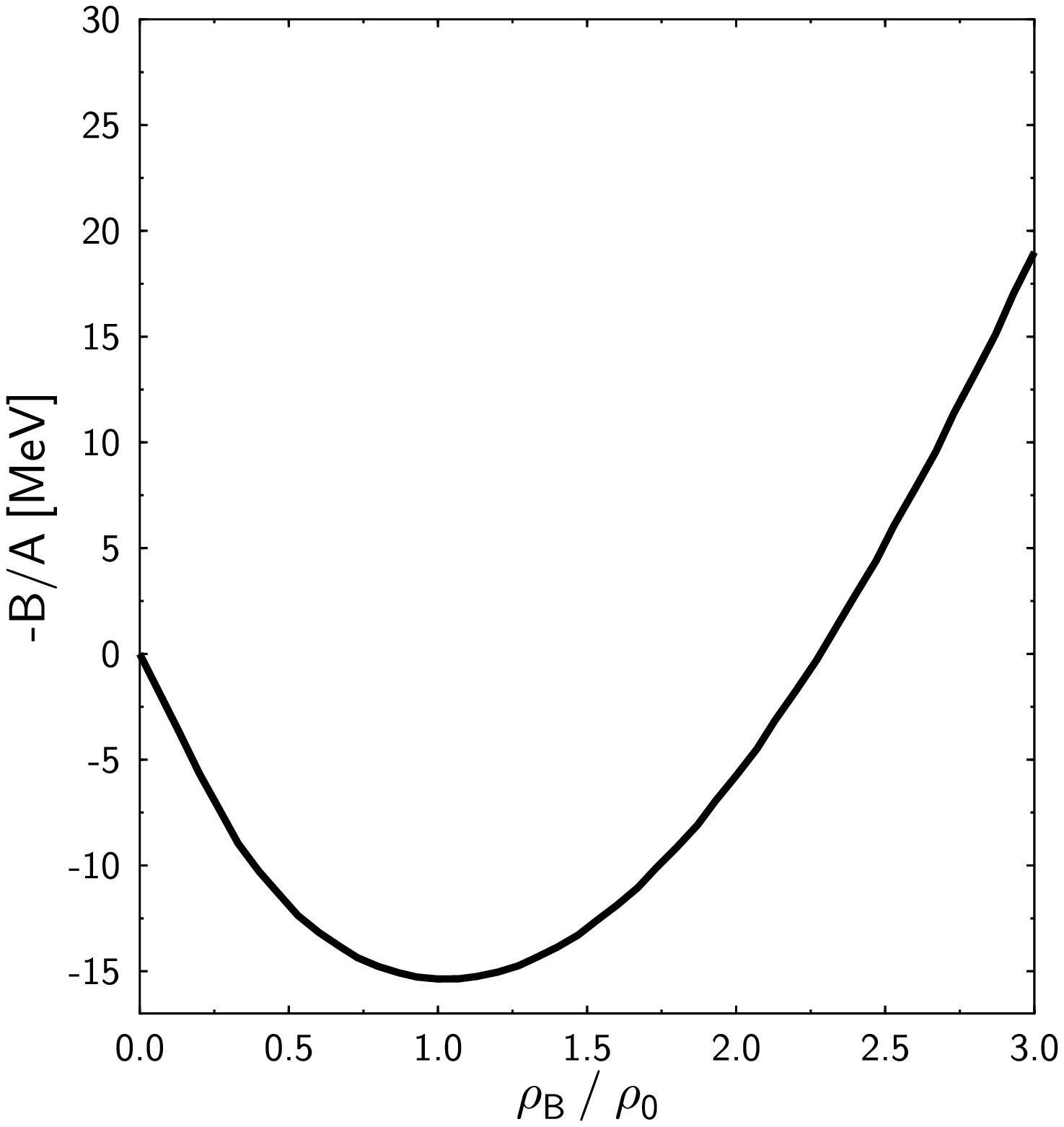}
\caption{\label{eos}Binding energy of nuclear matter as function of density. Right figure:
symmetric nuclear matter. Left figure: energy for varying isospin density
$\rho_p-\rho_n$ at fixed nuclear matter saturation
density $\rho=.153$ GeV/fm$^3$. Results with (full line)
and without (dotted curve) $\delta$ meson contributions are shown.
}
\end{figure}

Here the nonlinear terms in the vector self-interaction (\ref{Lvec}) and of the
scalar $\delta$ meson are crucial for a low value of
$E_{asym}$. In Fig. \ref{eos} one sees the influence of the $\delta$ meson on the asymmetry
energy. One can observe a softer energy dependence of the system on changing the isospin .
The total effect is a lowering of the asymmetry energy from $\sim 34.5$ to $31.9~$MeV.
The new parameter fit and the
addition of the $\delta$ field also has some effect on neutron star properties due
to the general softening of the nuclear equation of state. Compared to
an earlier calculation \cite{Nstar} the maximum mass of a (non-rotating) neutron star in this
model, determined by integrating the Tolman-Oppenheimer-Volkov equations as explained in
detail in \cite{Nstar}, is reduced to $M \sim 1.46~M_{solar}$ compared to $M \sim 1.64~M_{solar}$ without
the inclusion
of the $\delta$ meson.

\begin{figure}

\includegraphics[width = 17cm]{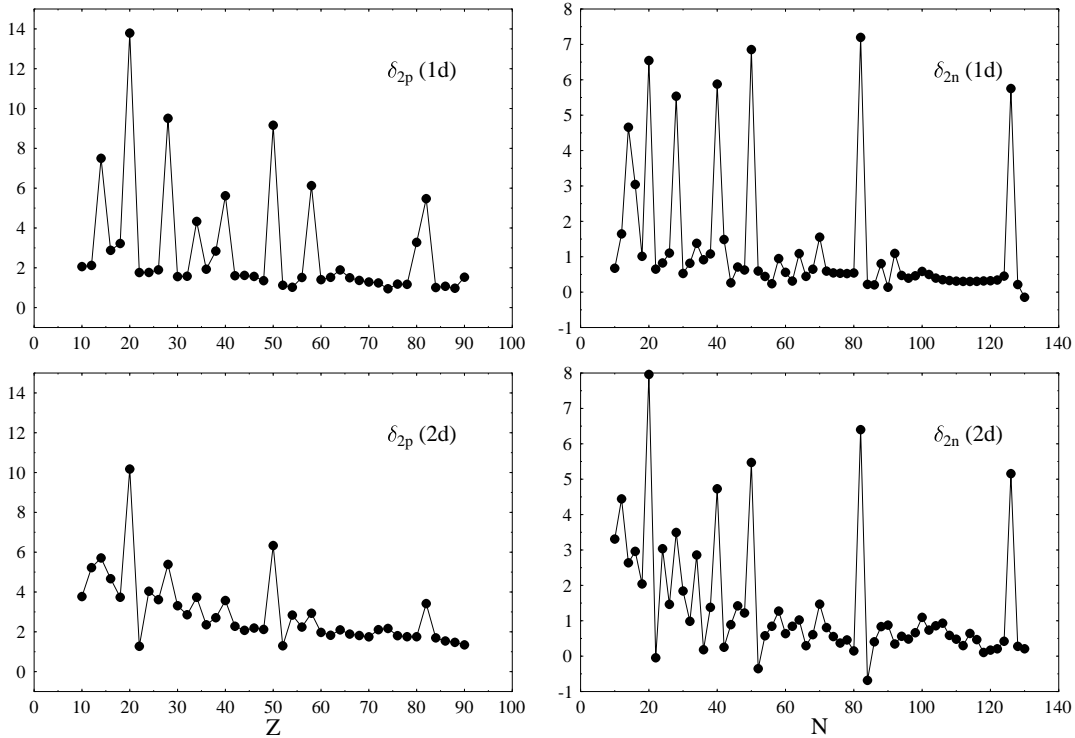}
\caption{\label{gap1d}Two-nucleon gaps $\delta_{2p}$ (left panels)
and $\delta_{2n}$  (right panels) in units of MeV for the $\chi_m$ parameter set.
The upper panels show the results of a calculation assuming spherical
symmetry.
The lower panels display the corresponding results for a 2-dimensional axially symmetric
calculation.
}
\end{figure}

We check the general applicability of our model by performing a full set of
calculations of 921 even-even nuclei (first in spherical approximation).
The upper panels of Fig.~\ref{gap1d} shows the result of a calculation of all even-even nuclei, listed in the
Audi-Wapstra-compilation \cite{audi} (augmented by possible superheavy nuclei up to a charge
$Z=136$, see Figure below). The two-proton and two-neutron gaps are displayed, defined as
\bea
\delta_{2p} (N,Z) &=& E(Z+2,N) - 2 E(Z,N) + E(Z-2,N)
\nonumber \\
\delta_{2n} (N,Z) &=& E(Z,N+2) - 2 E(Z,N) + E(Z,N-2)
\eea
In the plots the average value $\delta_{2p}(Z)$ of $\delta_{2p}(N,Z)$ for the various isotopes is shown:
\be
\delta_{2p}(Z) \equiv \frac{1}{n}\sum_{N} \delta_{2p}(N,Z)
\ee
where $n$ is the number of (even-even) isotopes. The same procedure is performed for the neutron gap:
\be
\delta_{2n}(N) \equiv \frac{1}{n}\sum_{Z} \delta_{2n}(N,Z)
\ee

\begin{figure}
\includegraphics[width = 15cm]{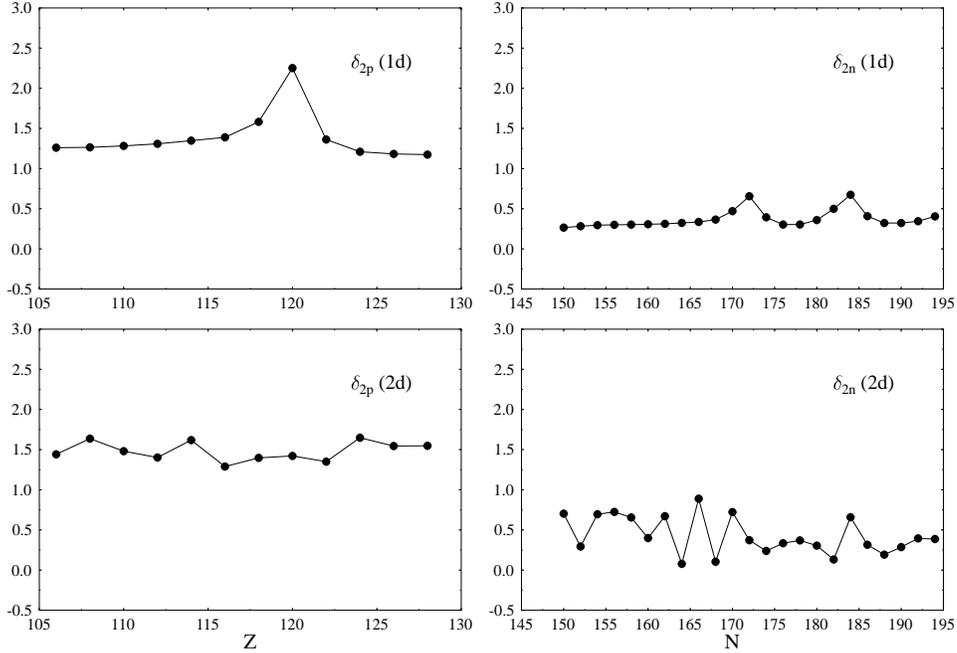}
\caption{\label{super1d}Two-nucleon gaps $\delta_{2p}$ (left panels)
and $\delta_{2n}$  (right panels) in units of MeV for the $\chi_m$ parameter set.
Results for hypothetical superheavy nuclei are shown.
The upper panels show the results of a calculation assuming spherical
symmetry.
The lower panels display the corresponding results for a 2-dimensional axially symmetric
calculation.
}
\end{figure}

One can clearly observe the major shell and subshell closures.
Looking beyond the tabulated nuclei in the range of the superheavies results for a spherical
calculation is shown in Fig. \ref{super1d}.
There is some signal for a shell gap that can be seen in the 1d calculation of $\delta_{2p}$ for
Z=120. There is no indication of a gap at the canonical value of Z=114.

Looking at $\delta_{2n}$ (upper right panel of Fig. \ref{super1d})
one can observe a weak signal for a shell closure at N=172 and N=184, which is in
accordance with most relativistic calculations (see \cite{bender}).
Note the strongly suppressed size of the gap compared to normal nuclei as seen in
Fig.~\ref{gap1d}.

Let us turn to first 2-dimensional calculations performed in this model.
Here the field equations are solved on a space-time grid adopting axial symmetry of
the system.
Part of the computer code is an adaptation of the RMF code by Rutz et al.\cite{Rutz}.
As pairing interaction we use a
zero-range pairing
interaction in the same way as it was already applied to RMF calculations \cite{bender2}.
The fitted pairing forces are quoted in Table \ref{table4}.
The resulting two-nucleon gaps in an axially symmetric 2-dimensional calculation are shown in the lower
panels of Fig.
\ref{gap1d}.
The general behavior is similar to the spherical results. The proton subshells are suppressed,
the major shell closures are clearly visible. The situation for neutrons is similar, there is
still a signal of a subshell at $N=40$.
If we look at the superheavy elements in the 2-dimensional calculation (lower panels in
Fig. \ref{super1d}
the signal is completely washed out, and at least the gap energy cannot serve
as a guideline for identifying shell closures. This originates from the strong deformation
of the nuclei involved in calculating equations (11). In the equation nuclei with
different deformations enter making it impossible to directly
infer conclusions on the shell structure of the nuclei.
This is similar to a possible shell quenching effect at Z=82 that is largely artificially
generated by comparing deformed nuclei above and below the shell closure as discussed
in \cite{bender}.

\begin{figure}

\includegraphics[width = 15cm]{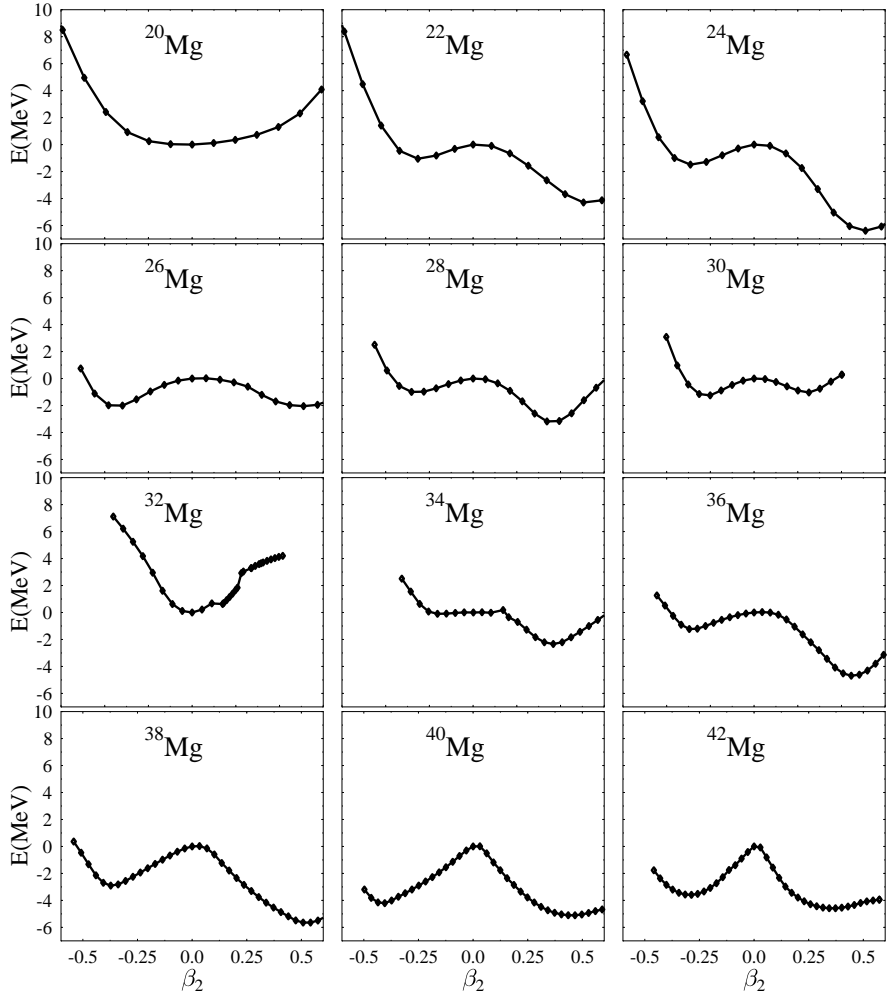}
\caption{\label{Mg}Energy (relative to the spherical calculation) of Mg isotopes as function of
quadrupole deformation $\beta_2$.}

\end{figure}

%
Comparing the 2d calculations with other theoretical calculations we study the chain of magnesium
isotopes. Here we perform a calculation with a constraint on the quadrupole deformation of
the nucleus (leaving all other multipoles unconstrained). We vary the constraint and calculate the energy of the isotopes as function
of their deformation. The resulting figure is shown in \ref{Mg}. These results that show the
occurrence of oblate and prolate minima for various isotopes is in agreement with other
relativistic and non-relativistic mean-field calculations \cite{Bproc,Buervenich, rod,laiz}.
There is no strongly deformed groundstate for $^{32}$Mg as experimental measurements of B(E2) values
suggest \cite{Mg32}.
One can, however, observe a shoulder in the energy at deformations $\beta_2 \sim 0.1$  to $0.2$, which
is due to an intruding $f7/2$ neutron state.

The overall quality of the model for  describing finite nuclei can be estimated
from the error in the binding energy averaged over all nuclei. The comparison has been done using
the Audi-Wapstra experimental data table. As a total average for nuclei with charge Z$=8$ and
higher we obtain a deviation $\epsilon_B \sim 0.35\%$ from the
the experimental binding energy values. Note, however, that most deviations arise from light
nuclei $A<=40$ where more subtle effects affect the binding energy that cannot accurately be
reproduced in a mean field-type description as applied here. Considering only heavier nuclei we
get the following numbers:
\begin{equation}
\epsilon^{\chi_m}_B (A>=50) = 0.21 \% , \quad \epsilon^{\chi_m}_B (A>=100) = 0.14 \%
\end{equation}
We can compare this result with an analogous calculation in the RMF model (using the NL3
parameter set with zero-range pairing force \cite{bender2})
\begin{equation}
\epsilon^{NL3}_B (A>=50) = 0.25 \% , \quad \epsilon^{NL3}_B (A>=100) = 0.16 \%
\end{equation}
We conclude that the quality of the nuclear fit within our model is slightly better quality compared to standard relativistic nuclear models.
Comparing the results with macroscopic-microscopic calculations of Moller and Nix, using their
data compilation \cite{MollerNix}, we read off values of $\epsilon^{MN}(A>=50) = 0.05\%$ and
$\epsilon^{MN}(A>=100) = 0.03\%$. This shows that in self-consistent models one is still off by a factor
of about 4 to 5 from those numbers (Note that in contrast to \cite{MollerNix}
the numbers from $\chi_m$ and NL3 arise from purely axially and time-symmetric calculations).

\begin{figure}

\includegraphics[width = 10cm]{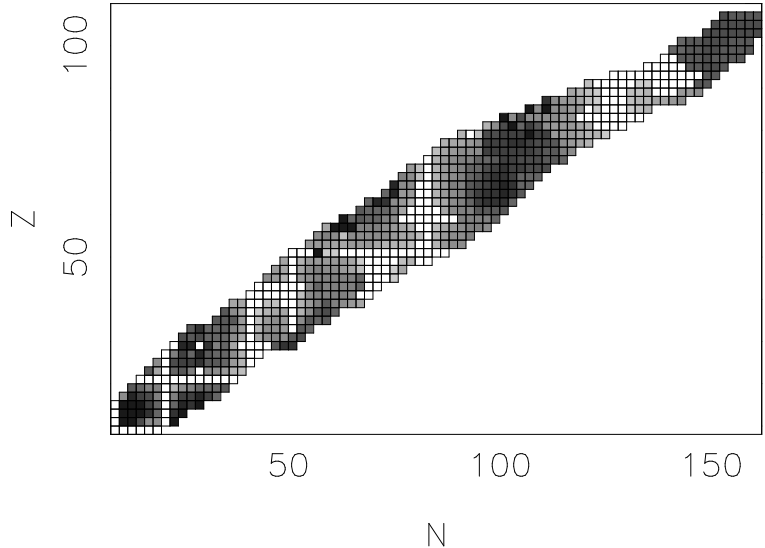}
\caption{\label{beta}Groundstate quadrupole deformation $|\beta_2|$ in our model using the $\chi_m$
parameters in a 2-dimensional calculation
assuming axially symmetry. White squares denote spherical shapes,
the darkest shade of gray represents deformations with $|\beta_2| >0.4$.
}
\end{figure}

We calculated the nuclear quadrupole deformations over the whole range of known nuclei assuming
axial symmetry. The calculation was done in the following way. We initialized each nucleus
starting with 3 configurations - prolate, oblate, and spherically symmetric and let the calculation
converge. We then took the deformation of the energetically lowest solution as groundstate
deformation of the nucleus. To a large extent this procedure reduces, but it does not completely eliminate,
the chance of missing out on the true minimum because
of shape isomers . There is still the possibility that a energetically lower state
has been missed. One can improve on this procedure by doing a constrained calculation for every nucleus,
varying the
quadrupole deformation and picking out the state with lowest energy. We will look at
some cases below. Obviously, this method needs much more computer time.  Here, the total time
used for generating Figure \ref{beta} was about 2500 CPU hours on a 500MHz Pentium machine.

\begin{figure*}
\includegraphics[width = 16cm, height=12cm ]{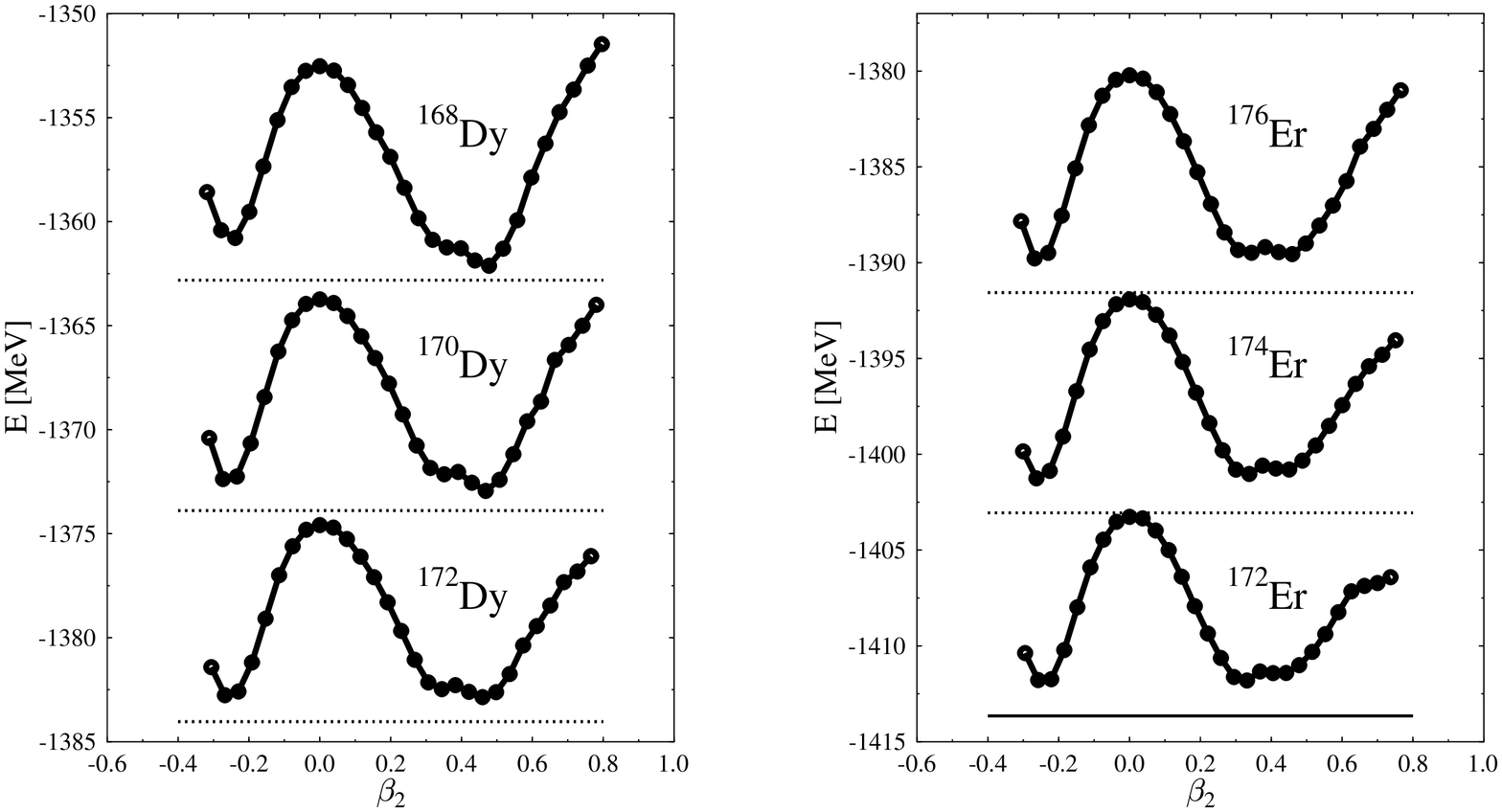}
\caption{\label{Dy172}Binding energies of Dy and Er isotopes, relative to the spherical state,
as function of deformation. Two nearly degenerate minima can be observed. The
horizontal lines show experimental (or extrapolated) values for the binding energy \cite{audi}
}
\end{figure*}

There are a number of areas where shape isomers exist. Especially in the $Z\sim70$ range and large
neutron number the calculated energies of the oblate and prolate minima are often within
a few hundred keV, and, given the approximations involved in this calculation, the true
groundstate deformation
cannot really be determined (in a more realistic calculation using configuration mixing both
minima will mix). It is possible that both states are connected via $\gamma$ deformation.
In future 3-dimensional calculations this point should be studied more closely.
As an example the Fig. \ref{Dy172} shows the resulting
deformation dependence of the energy of very neutron-rich Dysprosium and Erbium isotopes.
You see quite deformed prolate and oblate
minima with $\beta_2 \sim 0.45$ and $-0.28$, respectively, that are essentially degenerate.
In principle, there are two values for the deformation to be displayed in Fig. \ref{beta}.

\begin{figure}

\includegraphics[width = 10cm]{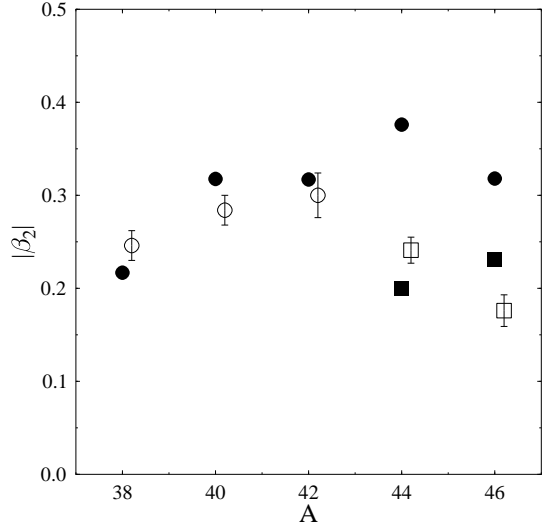}
\caption{\label{sulfur}Quadrupole deformation $|\beta_2|$ for sulfur (circles) and argon (squares)
isotopes with mass number A. The full symbols denote the theoretical results, the open symbols are experimental
data \cite{sulfur}.}
\end{figure}

\begin{figure}

\includegraphics[width = 10cm]{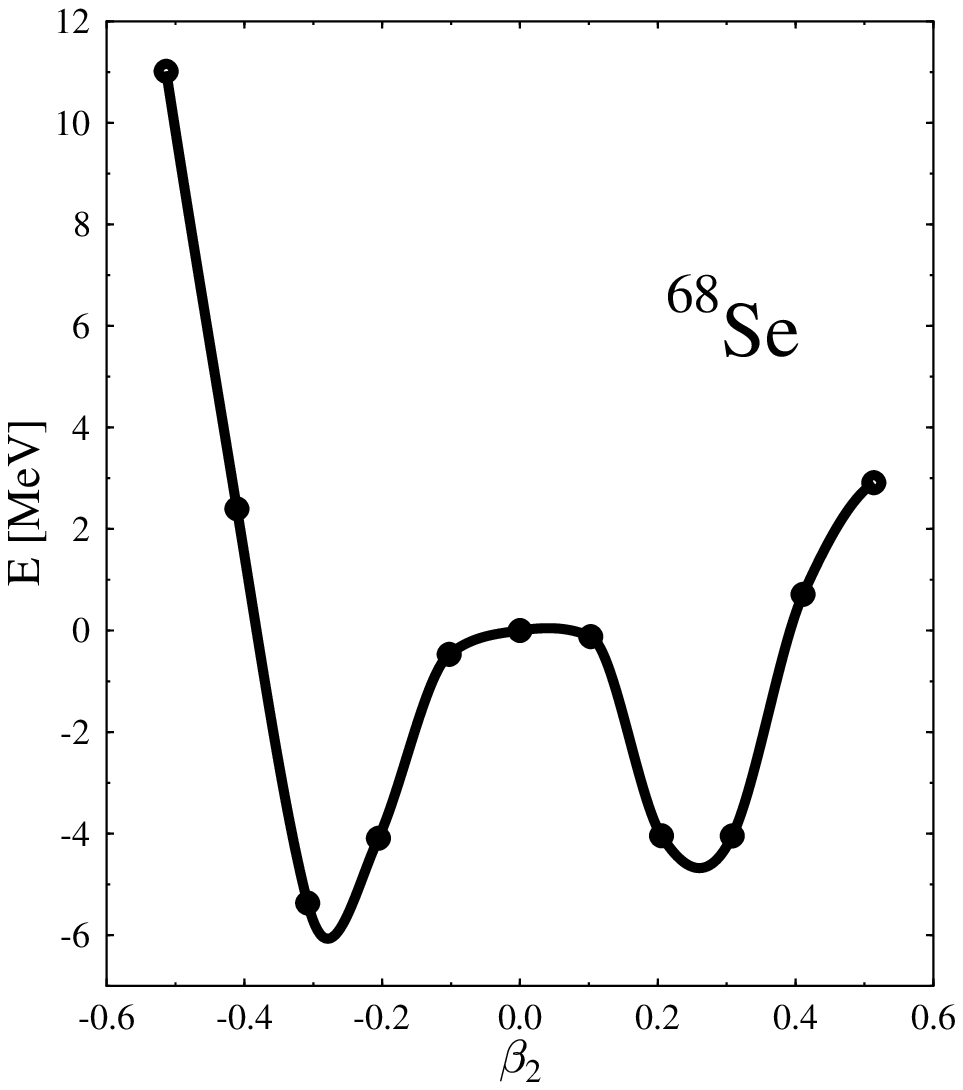}
\caption{\label{Se}Binding energy of the nucleus $^{68}$Se (relative to the spherical solution)
as function of the quadrupole constraint $\beta_2$.}
\end{figure}

\begin{figure}

\includegraphics[width = 16cm]{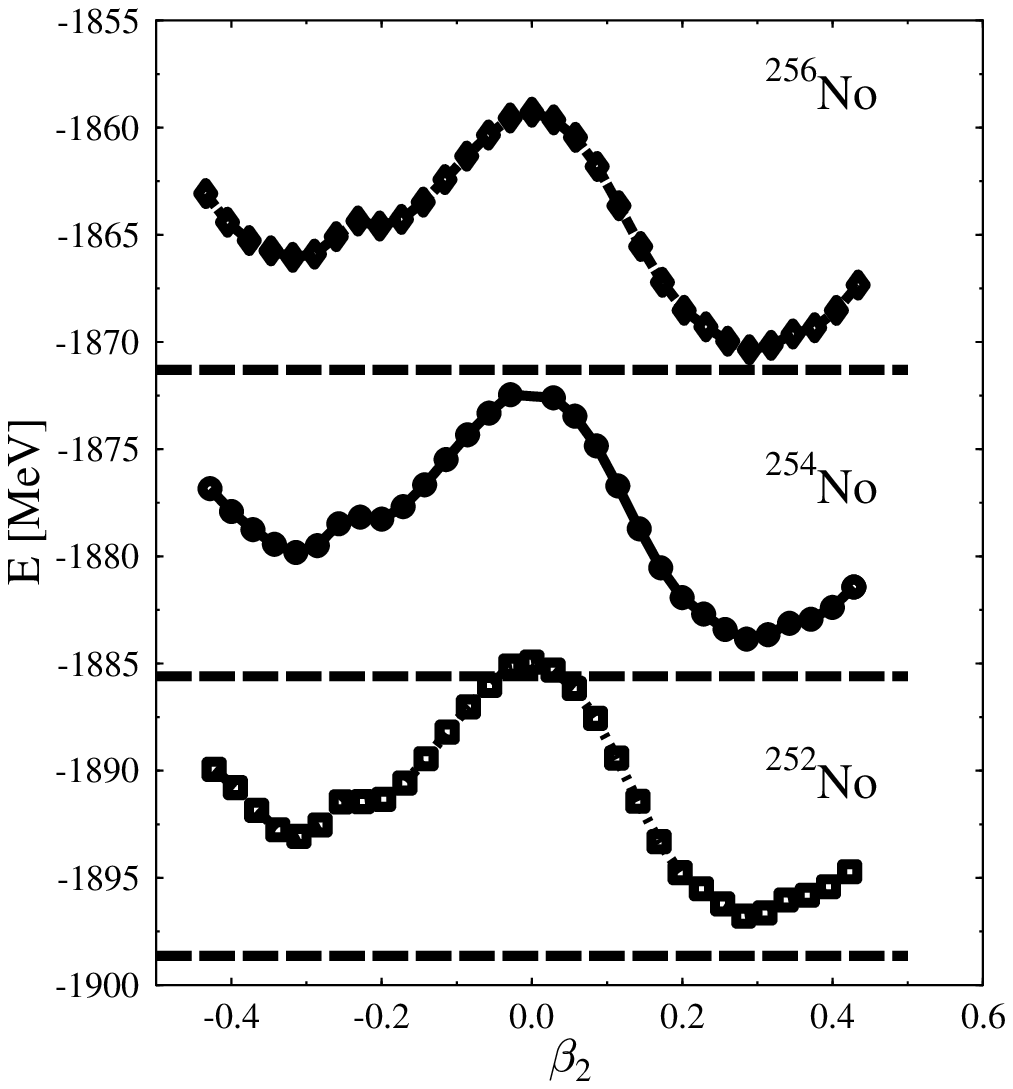}
\caption{\label{Nobelium}Binding energies of different Nobelium isotopes
as function of the quadrupole constraint $\beta_2$. The horizontal lines indicate the
measured groundstate energy}
\end{figure}

\begin{figure}
\includegraphics[width = 11cm,height=16cm]{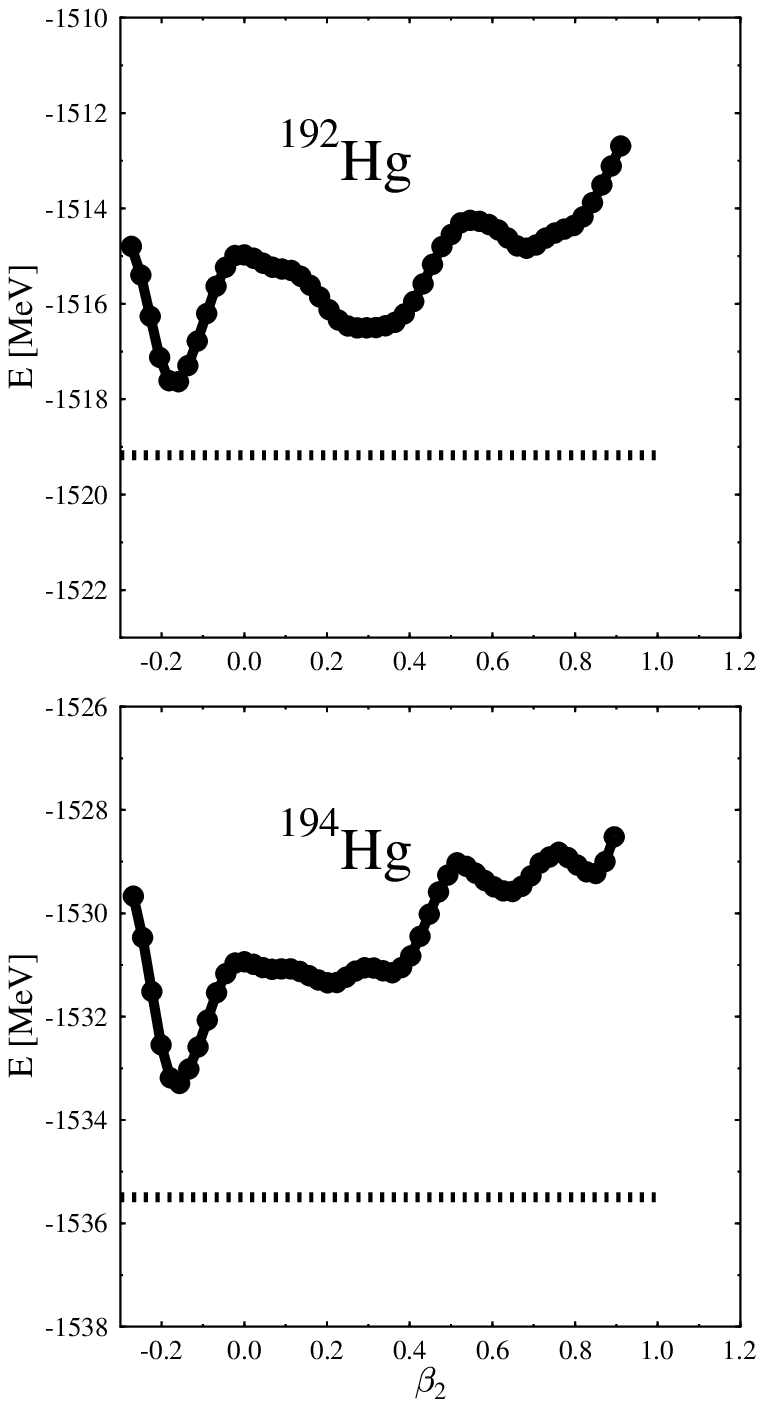}
\caption{\label{Hg}Binding energies of $^{192}$Hg and $^{194}$Hg as function of $\beta_2$.
Both isotopes exhibit an oblate groundstate and superdeformed structures for
$\beta_2 \sim 0.7$. Experimental binding energies are indicated by horizontal lines.}
\end{figure}

In a comparison with recent experimental results of the deformation
of various sulfur and argon isotopes \cite{sulfur} we calculated the groundstate
deformation of $^{38-46}$S and $^{42,44}$Ar. The results can be seen in Figure \ref{sulfur}.
The experiments determine the absolute value for $\beta_2$, the calculations show
prolate groundstates for sulfur and oblate ones for argon.
Our numbers agree quite well with experiment, some predictions for the sofar undetermined
deformation of very neutron rich sulfur isotopes are shown, too.
Going to proton-rich side we compare our results of a calculation of axially deformed
N=Z nucleus $^{68}$Se with relatively recent experimental data that show a strongly
oblate groundstate ($\beta_2 \sim -0.3$) together with an excited prolate state \cite{Fischer}.
The result of a quadrupole-constrained calculation is shown in Fig. \ref{Se}. In accordance
with the experiments distinct strongly deformed oblate and prolate minima are found, where
the oblate state is more strongly bound by about 2 MeV.
As a further test of calculations of deformed nuclei we look at
Nobelium isotopes. The calculation of the deformation of $^{252}$No,
$^{254}$No, and $^{256}$No are shown in Figure \ref{Nobelium}. The behavior of the
different isotopes is very similar, showing a prolate minimum at around $\beta_2 \sim 0.29$.
This compares very well with measurements \cite{No} of the groundstate deformation
of $^{252}$No and $^{254}$No with resulting numbers of $\beta_2 = 0.31 \pm 0.02$
for A=252 and $\beta_2 = 0.32 \pm 0.02$ for A = 254, respectively.
Finally, we investigated various superdeformed heavy nuclei. In all cases where a superdeformed
state is known we could either see a distinct superdeformed
minimum or a shoulder indicating the precursor of a superdeformed state for higher spins.
As an example Figure \ref{Hg} shows the result for the mercury isotopes $^{192}$Hg and $^{194}$Hg,
which has originally been predicted to have superdeformed bands in \cite{Chas}
(a study analogous to our discussion using in the RMF approach has be performed in \cite{Ring}).
\begin{table}
\caption{\label{table2}Parameter set $\chi_m$, all numbers are quoted for the appropriate powers
in MeV. $G_p$ and $G_n$ are the strengths of the zero-range pairing force for protons and
neutrons, respectively.}
\begin{ruledtabular}
\begin{tabular}{cccccccc}
$g_{N\sigma}$ & -10.569 &
$g_{N\zeta}$ & 0.467 &
$g_{N\omega}$ &  13.3265 &
$g_{N\rho}$ & 5.48851 \\
$k_0$ & 2.47584 &
$k_1$ & 1.35436 &
$k_2$ & -4.93719 &
$k_3$ & -2.77257 \\
$k_4$ & -0.233064 &
$g_4$ & 79.9083 &
$\epsilon$ & 0.02134&
$\chi_0$ & 409.76956\\
$G_p$ & $-4.47\cdot 10^{-5}$&
$G_n$ & $-4.34\cdot 10^{-5}$ &
$f_K$ & 122
\end{tabular}
\end{ruledtabular}
\end{table}
\begin{table}
\caption{\label{table4}Hadron masses (in MeV) for parameter set $\chi_m$. In brackets the
experimental isospin-averaged values are shown \cite{PDG}. Note that in the particle
data book $\sigma, \zeta, \delta$ correspond to $f_0(400-1200)$, $f_0(980)$, and $a_0(980)$,
respectively.}
\begin{ruledtabular}
\begin{tabular}{cccccccccccc}
$m_N$ & 939.2 &(938.9)&
$m_\Lambda$ & 1115.7 & (1115.6) &
$m_\Sigma$ & 1196.0 & (1193.1) &
$m_\Xi$ & 1331.3 & (1318.1) \\
$m_\pi$ & 138.6 & (138.0) &
$m_K$ & 497.3 & (495.6) &
$m_\eta$ & 574.0 & (547.5) &
$m_{\eta'}$ & 897.3 & (957.8) \\
$m_\sigma$ & 466.5 &(400 - 1200)&
$m_\zeta$ & 1024.5&(980)&
$m_\delta$ & 973.26 &(985)&
$m_\phi$ & 1020.0 &(1019.4)\\
$m_\omega$ & 780.6&(782.0)&
$m_\rho$ & 761.1&(768.1)&
$m_\chi$ & 960.0
\end{tabular}
\end{ruledtabular}
\end{table}
In addition to the oblate groundstate we see a rather complex structure
of the energy surface and clear evidence for superdeformation in both cases.
An experimental measurement and extrapolation to the groundstate of $^{194}$Hg gave an estimate of an
excitation energy $E \sim 6.02~$MeV \cite{Khoo} of the superdeformed state with respect to the oblate minimum.
In our calculation we get $E = 4.3~$MeV, which is a little bit low but
in reasonable agreement with experiment.
In the case of $^{192}$Hg we obtain $E=3.1~$MeV.

\section{Conclusions}
We investigated a hadronic model based on chirally symmetric
interactions within an flavor-SU(3) approach and calculated
nuclear properties within this approach. First we introduced a new
improved set of model parameters generated by a fit of a set of
spherical nuclei. For the first time also the isospin-triplet
$\delta$ meson was taken into account, which naturally occurs
within the SU(3) scheme in its coupling to the baryons as well as
in the nonlinear meson-meson interactions. The quality of the fit
is as good as the results of other relativistic mean field models
of nuclear structure. We investigated deformed nuclei and found
reasonable results for nuclear quadrupole deformations. A number
of improvements should be addressed in future calculations. A
projection onto good particle numbers as well as, in case of the
deformed calculations, a projection onto good angular momentum
should be performed. In different regimes of the nuclear landscape
shape isomers are found with  minima that are very close in their
energies. Especially for those cases three-dimensional studies and
a calculation including configuration mixing of the various states
have to be done. All these points are currently under
investigation \cite{later}.

\begin{acknowledgments}
The author wants to thank T. B\"urvenich and R. R. Chasman for helpful discussions and C. Rutz
for the use of parts of his nuclear mean-field code. This work
was supported by the U.S. Department of Energy, Nuclear Physics
Division (Contract No. W-31-109-Eng-38).
\end{acknowledgments}

\end{document}